\documentclass[a4paper,12pt]{article}
\usepackage{graphicx, rotating}
\usepackage{ifpdf}
\ifpdf
\usepackage{hyperref, pdfsync, epstopdf}    % This is for pdftex
\else
\usepackage[dvips,bookmarks]{hyperref}  % This is for arXiv.org
\fi
\hypersetup{colorlinks,bookmarksopen,bookmarksnumbered,citecolor=verdes,
linkcolor=blus,pdfstartview=FitH,urlcolor=rossos}
\def\hhref#1{\href{http://arxiv.org/abs/#1}{#1}} % in bibliography
\usepackage{amsmath}
\usepackage{slashed}

\renewcommand{\theequation}{\thesection.\arabic{equation}}

\newcommand{\beq}{\begin{equation}}
\newcommand{\eeq}{\end{equation}}
\newcommand{\fig}[1]{~\ref{fig:#1}}

\newcount\Mac  \Mac=1  % devo mettere Mac=1 se sto lavorando sul file Mac
\newcommand{\ifMac}[2]{\ifnum\Mac=1 #1 \else #2 \fi}
\def\putps(#1,#2)(#3,#4)#5#6{\ifnum\Mac=1 \put(#1,#2){\special{picture #5}}
\else  \put(#3,#4){\includegraphics{#6}} \fi}

\newcommand{\One}{\hbox{1\kern-.24em I}}

\renewcommand{\Im}{\mathop{\rm Im}}

\newcommand{\GeV}{\,{\rm GeV}}

\newcommand{\PRL}{Phys. Rev. Lett.}

\newcommand{\PR}{Phys. Rev.}
\newcommand{\eq}[1]{~{\rm (\ref{eq:#1})}}

\renewcommand{\Re}{\hbox{Re}\,}

\newcommand{\lascia}[1]{}
\makeatletter
\def\art{\@ifnextchar[{\eart}{\oart}}
\def\eart[#1]#2#3#4#5#6{{\rm #2}, {#3 #4} {\rm (#6) #5} [{\hhref{#1}}]}
\def\hepart[#1]#2{{\rm #2, \hhref{#1}}}
\newcommand{\oart}[5]{{\rm #1}, {#2 #3} {\rm (#5) #4}}

\newcounter{alphaequation}[equation]
\def\thealphaequation{\theequation\hbox to
0.6em{\hfil\alph{alphaequation}\hfil}}
\def\eqnsystem#1{
\def\@eqnnum{{\rm (\thealphaequation)}}
\def\@@eqncr{\let\@tempa\relax \ifcase\@eqcnt \def\@tempa{& & &} \or
  \def\@tempa{& &}\or \def\@tempa{&}\fi\@tempa
  \if@eqnsw\@eqnnum\refstepcounter{alphaequation}\fi
\global\@eqnswtrue\global\@eqcnt=0\cr}
\refstepcounter{equation} \let\@currentlabel\theequation \def\@tempb{#1}
\ifx\@tempb\empty\else\label{#1}\fi
\refstepcounter{alphaequation}
\let\@currentlabel\thealphaequation
\global\@eqnswtrue\global\@eqcnt=0 \tabskip\@centering\let\\=\@eqncr
$$\halign to \displaywidth\bgroup \@eqnsel\hskip\@centering
$\displaystyle\tabskip\z@{##}$&\global\@eqcnt\@ne
\hskip2\arraycolsep\hfil${##}$\hfil& \global\@eqcnt\tw@\hskip2\arraycolsep
$\displaystyle\tabskip\z@{##}$\hfil
\tabskip\@centering&\llap{##}\tabskip\z@\cr}

\def\endeqnsystem{\@@eqncr\egroup$$\global\@ignoretrue} \makeatother

\newcommand{\mb}[1]{\mbox{\normalsize\boldmath $#1$}}

\def\Lag{{\cal L}}
\def\SU{{\rm SU}}

\def\circa#1{\,\raise.3ex\hbox{$#1$\kern-.75em\lower1ex\hbox{$\sim$}}\,}

\usepackage{multicol}
\usepackage{color}
\definecolor{rosso}{cmyk}{0,1,1,0.4}
\definecolor{rossos}{cmyk}{0,1,1,0.55}
\definecolor{rossoc}{cmyk}{0,1,1,0.2}
\definecolor{blu}{cmyk}{1,1,0,0.3}
\definecolor{blus}{cmyk}{1,1,0,0.6}
\definecolor{bluc}{cmyk}{1,1,0,0.1}
\definecolor{verde}{cmyk}{0.92,0,0.59,0.25}
\definecolor{verdec}{cmyk}{0.92,0,0.59,0.15}
\definecolor{verdes}{cmyk}{0.92,0,0.59,0.4}
\definecolor{grigio}{cmyk}{0,0,0,0.07}
\definecolor{rosa}{cmyk}{0,0.1,0.1,0.02}
\definecolor{rosino}{cmyk}{0,0.05,0.05,0.02}
\definecolor{rosas}{cmyk}{0,0.3,0.25,0.05}
\definecolor{celeste}{cmyk}{0.1,0,0,0.02}
\definecolor{giallino}{cmyk}{0,0,0.4,0.02}
\definecolor{rosso}{cmyk}{0,1,1,0.4}
\definecolor{rossos}{cmyk}{0,1,1,0.55}
\definecolor{rossoc}{cmyk}{0,1,1,0.2}
\definecolor{blu}{cmyk}{1,1,0,0.3}
\definecolor{bluc}{cmyk}{1,1,0,0.1}
\definecolor{blucc}{cmyk}{0.7,0.5,0,0}
\definecolor{viola}{cmyk}{0,1,0,0.6}
\definecolor{viola2}{cmyk}{0,1,0.2,0.6}
\definecolor{verde}{cmyk}{0.92,0,0.59,0.25}
\definecolor{verdec}{cmyk}{0.92,0,0.59,0.15}
\definecolor{verdes}{cmyk}{0.92,0,0.59,0.4}
\definecolor{verdino}{cmyk}{0.12,0,0.09,0.05}
\definecolor{giallo}{cmyk}{0,0,1,0}
\definecolor{gialloverde}{cmyk}{0.44,0,0.74,0}

\font\tenrsfs=rsfs10 at 12pt

\font\sevenrsfs=rsfs7
\font\fiversfs=rsfs5
\newfam\rsfsfam
\textfont\rsfsfam=\tenrsfs
\scriptfont\rsfsfam=\sevenrsfs
\scriptscriptfont\rsfsfam=\fiversfs
\def\mathscr#1{{\fam\rsfsfam\relax#1}}
\def\Lag{\mathscr{L}}

\begin{document}

\vspace{-4cm}
\hspace{8cm}
FTUAM-13-29, \, \,
IFT-UAM/CSIC-13-113

\color{black}
\vspace{0.5cm}
\begin{center}
{\Huge\bf\color{rossos}Thermal axion production}\\
\bigskip\color{black}\vspace{0.6cm}{
{\large\bf  Alberto Salvio,}$^a$
{\large\bf Alessandro Strumia}$^{b,c}$ {\large and \bf Wei Xue}$^d$
} \\[7mm]
{\it $^a$ Departamento de F\'isica Te\'orica, Universidad Aut\'onoma de Madrid and \\ 
Instituto de F\'isica Te\'orica IFT-UAM/CSIC, Cantoblanco, 28049 Madrid, Spain}\\[3mm]
{\it $^b$ Dipartimento di Fisica dell'Universit{\`a} di Pisa and INFN, Italia}\\[3mm]
{\it  $^c$ National Institute of Chemical Physics and Biophysics, Tallinn, Estonia}\\[3mm]
{\it $^d$ INFN, Sezione di Trieste, SISSA, via Bonomea 265, 34136 Trieste, Italy}\\[3mm]
\end{center}
\bigskip
\centerline{\large\bf\color{blus} Abstract}
\begin{quote}\large
We reconsider thermal production of axions in the early universe, including  axion couplings
to all Standard Model (SM) particles.
Concerning the axion coupling to gluons, we find
that thermal effects enhance the axion production rate by a factor of few with respect to
previous computations performed in the limit of small strong gauge coupling.
Furthermore, we find that the top Yukawa coupling induces a much larger axion production rate,
unless the axion couples to SM particles only via anomalies.
\color{black}
\end{quote}
{\small\tableofcontents}

\newpage

\section{Introduction}
The strong CP problem can be solved by a Peccei-Quinn (PQ) symmetry~\cite{La}, that manifests at low energy as a light axion $a$.
The axion is a good dark matter (DM) candidate, if cold axions are produced non-thermally via the initial misalignment mechanism~\cite{axionDM}.
The cosmological DM abundance is reproduced for an order one initial misalignment angle provided that $f_a\approx 10^{11}\GeV$,
which is compatible with the experimental bound $f_a\circa{>} 5\times 10^9\GeV$~\cite{KSVZ,DFSZ}.
The ADMX experiment can probe such scenario in the next years~\cite{ADMX}.

\begin{figure}[t]
\begin{center}
$$\includegraphics[width=0.6\textwidth]{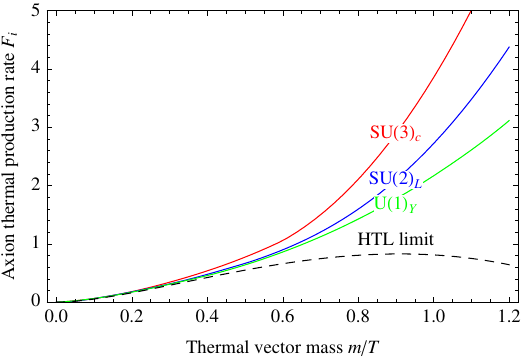}$$
\caption{\label{fig:res}\em 
Our result for the axion production rate as function of the thermal mass of vectors.
The functions $F_{1,2,3}(m/T)$ are defined in eq.~(\ref{defF321}) and the thermal masses
of the vectors within the SM are 
$m/T=g_3$ for  gluons, $m/T=\sqrt{11/12} g_2$ for the $W,Z$ and $m/T = \sqrt{11/12} g_Y$ for hypercharge.
For comparison, the lower dashed curve is the 
result of~\cite{Graf:2010tv} (eq.~(\ref{HTLresutl}))
computed within the HTL approximation, valid in the limit $g_3\ll1$.
%IF THE FOLLOWING SENTENCES ARE REMOVED A STRANGE LATEX BUG APPEARS
%a) summing the Feynman diagrams in the upper row, 
% the diagrams relative to a simplified world without quarks.
}
\end{center}
\end{figure}

Furthermore, thermal scatterings in the early universe unavoidably produce a population of hot axions.
The goal of this paper is performing an improved computation of hot axion thermal production.
The thermal axion production  rate~\cite{Masso} was previously computed
in~\cite{Graf:2010tv} making use of the Hard Thermal Loop (HTL)~\cite{Pisarski92,LeBellac} approximation ($g_3\ll 1$),
and considering only the axion coupling to gluons.
The resulting space-time density of thermal axion production was:
\beq
\gamma_{a}^{\rm HTL} =\frac{\zeta(3)g_3^4T^6}{64\pi^7f_a^2} F_3^{\rm HTL},\qquad
 F_3^{\rm HTL}= g_3^2 \ln \frac{1.5^2}{g_3^2} . \label{HTLresutl}\eeq
However, this HTL production rate unphysically decreases for $g_3\circa{>} 1.3$ becoming
negative for $g_3\circa{>}1.5$.
Fig.\fig{res} shows that the physical value, $g_3\approx 0.85$ at $T\sim 10^{10}\GeV$,
lies in the region where the HTL rate function $F_3^{\rm HTL}(g_3)$ (lower dashed line) is unreliable. 
Fig.\fig{res} also illustrates our final result: as described in the following sections,
$F_3^{\rm HTL}$ will be replaced by $F_3$, plotted in the upper line,
which agrees with the HTL result in the limit $g_3  \ll 1$,
and is about twice larger for the physical value of $g_3$.

%The basic point of our improved computation is including the contribution of $g_T \to g_T a$ `decays'.
%Here $g_T$ denotes a gluon excitation of the thermal plasma.
%Such excitations (that in the zero temperature limit become the usual gluons)
%have a non-relativistic dispersion relation
%(because the thermal plasma sets a preferred frame)
%and a non-vanishing spectral density both above and below the light cone.
%At lowest order in $g_3$ such `thermal decay' is equivalent to some $2\to 2$ scatterings,
%but its full computation allows us to include those higher order effects which are enhanced by the $\sim (4\pi)^2$
%factor characteristic of $1\to 2$ phase space with respect to $2\to 2$ phase space.

%To get the axion production rate we actually
%compute the imaginary part of the axion propagator in the
%thermal plasma by making use of the full resummation techniques for the internal propagators at finite temperature. Thermal effects distort the dispersion relations
%$E(k)$ of gluons and quarks by i) adding a thermal
%mass $E^2 = k^2 + m^2(k)$ to the modes already existing at  zero
%temperature; ii) by introducing  new collective excitations (gluons
%with longitudinal polarization, etc)
%with their own dispersion relation; iii) beyond the two poles
%mentioned above, the spectral densities of particles in a thermal
%plasma also develop a `continuum' contribution, that can be thought
%of as a parton-like distribution, with a continuum range of masses.
%Physically it arises because particles can exchange energy with the
%plasma.

Furthermore, we go beyond the anomalous axion coupling to gluons (a loop level effect),
computing the axion production rate due to all axion couplings.
We  find that the axion coupling to top quarks (a tree level effect)
gives an axion production rate which is about 3 orders of magnitude larger,
unless the axion couples to SM particles only via anomalies
%\footnote{\xxx{This is, however, possible: it is for example realized in  in KSVZ axion models~\cite{KSVZ}, where the SM fields have vanishing PQ charges.}}.

In section~\ref{subtractions} we outline our computation, performed
in section~\ref{subtracted scattering rate} (subtracted scattering rates)
and in section~\ref{Full axion  production rate} (higher order enhanced effects).
In section~\ref{res} we present  our final result and discuss its cosmological implications.
Conclusions are presented in section~\ref{concl}.
An off-topic but important subtlety is discussed in a footnote.\footnote{We show that scattering
involving many particles can be neglected.
This is trivially true in quantum field theory: 
for example the cross section for $2\to 3$ scatterings is $g^2/(4\pi)^2$ times smaller than the cross section of the dominant $2\to 2$ scatterings.
However, this is not generically true in thermal field theory, where the expansion parameter is $g$
(rather than $g^2/(4\pi)^2$) and where 
collinear kinematical configurations
 can  enhance higher order scatterings by powers of $1/g$, such that a resummation of $2\to n$ scattering becomes needed.
This subtlety was noticed in the context of computations of photon emission from a quark-gluon plasma~\cite{BY}:
for example, the $2\to 3$ process constructed adding to a $qg\to qg$ scattering a
a $q\to q \gamma$ vertex, where $q$ and $\gamma$ are almost collinear
(the directions of their moment differ by a small angle $\theta$) allows a kinematical configuration
where the propagator of the virtual gluon that mediates the scattering is enhanced by $1/\theta^2$,
while the gauge vertex $q\to q\gamma$ is only suppressed by $\theta$.  
This results into a $1/\theta$ enhancement of the scattering amplitude, cut-off by the thermal mass $m\sim gT$,
and thereby to a $1/g^2$ enhancement of the $2\to 3$ scattering rate.

We verified that no such collinear enhancement is present for axion production, because the axion vertices
(such as the axion/gluon/gluon vertex $a G_{\mu\nu}\tilde{G}_{\mu\nu}$) are suppressed as $\theta^2$ in the collinear limit.
We also verified that the similar vertices relevant for graviton, gravitino~\cite{Rychkov:2007uq}, axino~\cite{axino} production
similarly lead to no collinear enhancement, being $\theta^2$ suppressed.
Thereby there is no need of adding higher order scatterings and previous computations remain valid.}

\medskip

%[

%In agreement with na\"{\i}ve dimensional analysis, the
%result for the number of scatterings per space-time volume at leading order in $g$
%has the form
%\beq \gamma_{\rm scattering} \sim\frac{T^6}{\pi^5 \bar M_{\rm Pl}^2}   g^2 \ln\frac{1}{g} \eeq
%where
%We now explain why this result is unsatisfactory.
%The practical reason can be easily seen looking at t

%

%\begin{figure}
%\begin{center}
%$$\includegraphics[width=0.6\textwidth]{figs/rate}$$
%\caption{\label{fig:res}\em Our result for the axion production rater
%as function of the strong coupling $g_3$
%(upper curve, the quantity $F(g_3)$ is defined in eq.~(\ref{F-def})) compared to the
%result of~\cite{Graf:2010tv} (lower dashed curve, eq.~(\ref{HTLresutl}))
%computed within the HTL approximation, valid in the limit $g_3\ll1$.}
%\end{center}
%\end{figure}

\setcounter{equation}{0}

\section{Outline of the computation}\label{subtractions}

\subsection{Effective axion Lagrangian}
The effective action that describes axion couplings to SM particles at first order in the axion field $a$ is written 
in the basis where
the SM Lagrangian $\Lag_{\rm SM}$ does not contain the axion as~\cite{La}
\begin{eqnarray} 
\Lag&=& \Lag_{\rm SM}+ \frac{(\partial_\mu a)^2}{2} -
\frac{a}{f_a}
\bigg[c_3\frac{\alpha_3}{8\pi}  G_{\mu\nu}^a \tilde G_{\mu\nu}^a+
c_2 \frac{\alpha_2}{8\pi}  W_{\mu\nu}^a \tilde W_{\mu\nu}^a+
c_1 \frac{\alpha_Y}{8\pi} B_{\mu\nu} \tilde B_{\mu\nu}\bigg]+\\
&&+
\frac{\partial_\mu a}{f_a}\bigg[  c_H   H^\dagger i (D_\mu H) 
-c_Hi(D_\mu H)^\dagger H \bigg]+\frac{\partial_\mu a}{f_a}\sum_\psi    c_\psi (\bar\psi \gamma_\mu \psi). 
\nonumber
\end{eqnarray}
Here $\tilde{G}_{\mu\nu} = \frac{1}{2} \varepsilon_{\mu\nu\alpha\beta} G_{\alpha \beta}$
are the field strength duals;  $\varepsilon_{0123}=1$;
$H$ is the Higgs doublet;
the Weyl spinors $\psi=\{Q,U,D,L,E\}$ are  the SM fermions;
$f_a$ is the effective axion decay constant in the convention where $c_3=1$;
$c_1$ and $c_2$ are the axion couplings to electro-weak vectors;
$c_H$ is the axion coupling to the Higgs;
$c_\psi$  are the axion derivative couplings to the SM fermions.
All $c$ coefficients are real and dimensionless.
In the full axion theory $c_H$ and $c_\psi$ are the PQ charges of the SM fields
(they vanish in KSVZ axion models~\cite{KSVZ}), while $c_{1,2,3}$ also receive contributions
from extra heavy fermions (not present in DFSZ axion models~\cite{DFSZ}).\footnote{As usual, the effective action above is reliable only at energies much below the masses of the extra non-SM fields present in the axion model one considers.
For example, if the KSVZ extra fermions with non-vanishing PQ charges were light enough to be present in the thermal bath,
they would give an extra contribution to the axion production rate at tree level,
that would dominate with respect to one-loop contribution that we consider, encoded in the anomaly coefficients $c_{1,2,3}$. 
The computation of such extra contribution would be analogous to the top Yukawa contribution discussed below.
}

While in previous computations only the axion/gluon coupling was considered,
we want to consider all axion couplings.

For this purpose, it is convenient to perform a phase redefinition of the SM matter fields
\beq \psi \to e^{i c_\psi a/f_a} \psi,\qquad H\to e^{i c_H a/f_a} H \label{eq:trans}\eeq
such that, at the first order in $a$, the $c_\psi$ and $c_H$ couplings are removed, at the price of shifting the axion coupling to vectors as follows
\beq \begin{array}{l}
c_3  \to c_3' \equiv c_3 +\sum(c_U +c_D- 2 c_Q ),\\
c_2 \to c_2'  \equiv c_2 -\sum(3 c_Q +c_L) ,\\
 c_1\rightarrow c_1'  \equiv c_1+\sum(2c_E-c_L+\frac{8}{3}c_U+\frac{2}{3}c_D-\frac{1}{3}c_Q),
\end{array}  \label{cred}
 \eeq
where the sum runs over the 3 fermion generations. 
We used the fact that all SM matter field lie in fundamental representations with generators $T^a$
normalized as  Tr$(T^aT^b)=\delta^{ab}/2$. 
The transformation\eq{trans} also introduces an axion phase in the SM Yukawa couplings.
For our purposes, all SM Yukawa couplings are negligibly small except  the top Yukawa, for which the transformation induces the following axion phase:
\beq y_t \rightarrow  y_t \exp\left[i c'_t \frac{a}{f_a}\right],\qquad
c'_t \equiv  c_H+c_{Q_3}-c_{U_3}  .
%=y_t \left[1+i(c_H+c_{Q_3}-c_{U_3})\frac{a}{f_a}\right] +O(a^2)
\eeq
So, at first order in $a$, the transformation generates the following Lagrangian interaction:
\begin{equation} i c'_t  y_t  \frac{a}{f_a} Q_3HU_3 +\hbox{h.c.} \, .   \label{lambdatprime}\end{equation}
The thermal axion production rate will be computed in terms of $c'_3$
(strong interactions), $c'_2$ (weak interactions), $c'_1$ (hypercharge) and
$c'_t$ (top Yukawa coupling).

%\footnote{
%Their values can be arbitrarily shifted $c_i \to c_i + \hbox{cte}\, Q_i$ by using the
%U(1) symmetries $Q$ of the SM (hypercharge, baryon number, lepton flavours)
%to globally redefine the phases of the SM fields.
%In general the phase of the SM fermion mass terms  depend on the axion; however
%the Lagrangian is invariant under shifts $a(x) \to a(x) + \hbox{constant}$, 
%and this invariance can be used to perform a global phase redefinition
%of the SM fermions such that the axion only has derivative couplings and 
%anomalous terms.}

%In this paper we mostly focus on the coupling of  the axion to gauge fields. We include, however, the coupling of the axion to fermions and to the Higgs field by using eqs. (\ref{cred})-(\ref{lambdatprime}).
%, $-a {\cal O}/f_a$ with
%${\cal O} = \frac{\alpha_3}{8\pi}  G_{\mu\nu}^a \tilde G_{\mu\nu}^a$.
% takin into account the coupling of the gluons to quarks.

\medskip

\begin{figure}[t]
\begin{center}
$$\includegraphics[width=0.9\textwidth]{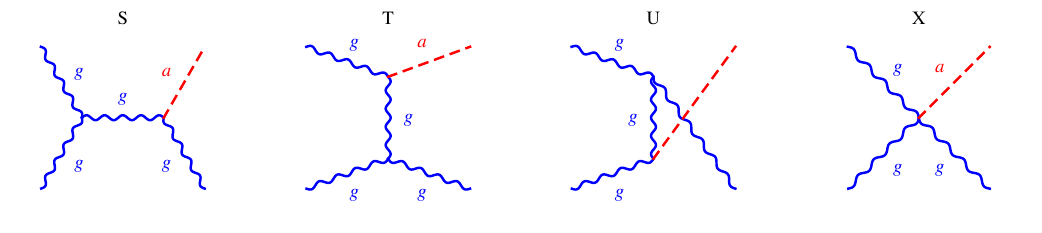}$$
$$\includegraphics[width=0.98\textwidth]{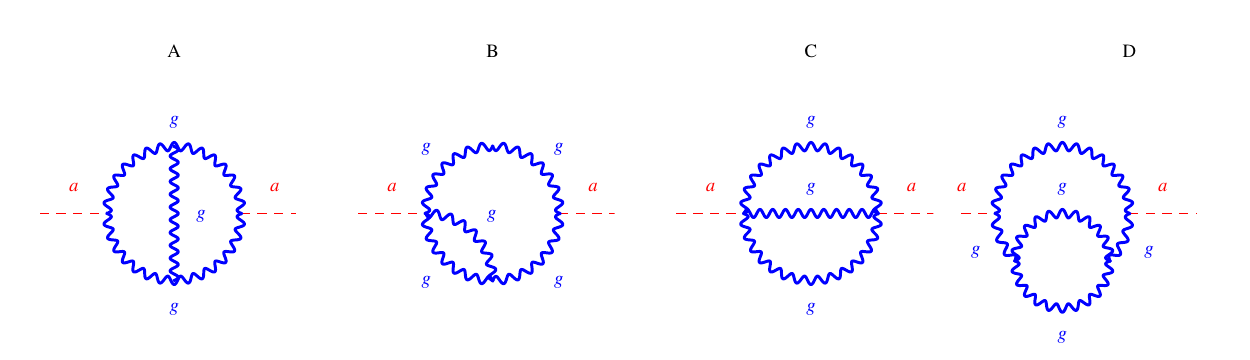}$$
\caption{\label{fig:FeynA}\em 
The leading-order $gg\to ga $ scattering rate in the thermal plasma 
is equivalently obtained by: 
a) summing the Feynman diagrams in the upper row, squaring the total amplitude, performing the thermal average;
b) summing the imaginary parts of the two loop thermal diagrams in the lower row.
In both cases the result is infra-red divergent, such that proper inclusion of higher order effects is needed.
For simplicity, we here plotted the diagrams relative to a simplified world without quarks.
}
\end{center}
\end{figure}

\subsection{Thermal production rate}
According to the general formalism of thermal field theory \cite{LeBellac}, the thermal production
rate of  a weakly interacting scalar $a$ is equivalently computed
from the
imaginary part of its propagator $\Pi_a$ as
\begin{equation}
\gamma_a = \frac{dN_a}{dV\,dt}=-2\int d\vec{P}\,f_{B}(E)\Im\Pi_a =\int d\vec{P}%
~\Pi^{<}_a(P),\qquad d\vec{P}\equiv\frac{d^{3}p}{2E(2\pi)^{3}} .
\label{eq:ImProp}%
\end{equation}
Here $\Pi^{<}_a$ is the non time-ordered axion propagator and $P = (E,\vec p)$.
Thermal field theory cutting rules allow to see that, at leading
order in the SM couplings, eq.\eq{ImProp} is equivalent to the usual summing of
all rates for the various tree-level processes that lead to axion
production. 

\smallskip

We  illustrate  the general discussion with the concrete example
of the axion coupled to a simplified SM consisting only of gluons.
In such a case the thermal axion production rate $\gamma$
at leading order in $g_3$ can be obtained
by computing the $gg\to ga$ scattering rate 
and thermally averaging it.
\begin{itemize}
\item The  Feynman diagrams for $gg\to ga$ scatterings
are plotted in the upper row of figure\fig{FeynA} and are named
$S$ ($s$-channel gluon exchange), $T$ ($t$-channel), $U$ ($u$-channel) and $X$ (quartic vertex).
When computing the rate in terms of scatterings, the rate is proportional to the modulus squared of the total
amplitude, $|S + T + U + X|^2$.

\item
The equivalent thermal diagrams at leading order in $g_3$ arise at two-loop level and
are plotted in the lower row of figure\fig{FeynA}, where they are named $A$, $B$, $C$, $D$.
The rate is proportional to the their sum,
that contains the various tree-level scatterings in the following way:
\beq \begin{array}{ll}
A=2 \Re [S^* T+S^*U+T^*U]
\qquad&
B= 2 \Re [X^* (S+T+U)],\\
C= |X|^2 \qquad &
D=|S|^2+ |T|^2 + |U|^2. 
\end{array}\label{eq:ABCD}
 \eeq
 \end{itemize}
We explicitly see that the thermal axion production rate $\gamma = \gamma_A + \gamma_B + \gamma_C + \gamma_D$ is equivalent to 
the scattering computation $|S+T+U+X|^2$.

\bigskip

\begin{figure}[t]
\begin{center}
$$\includegraphics[width=0.98\textwidth]{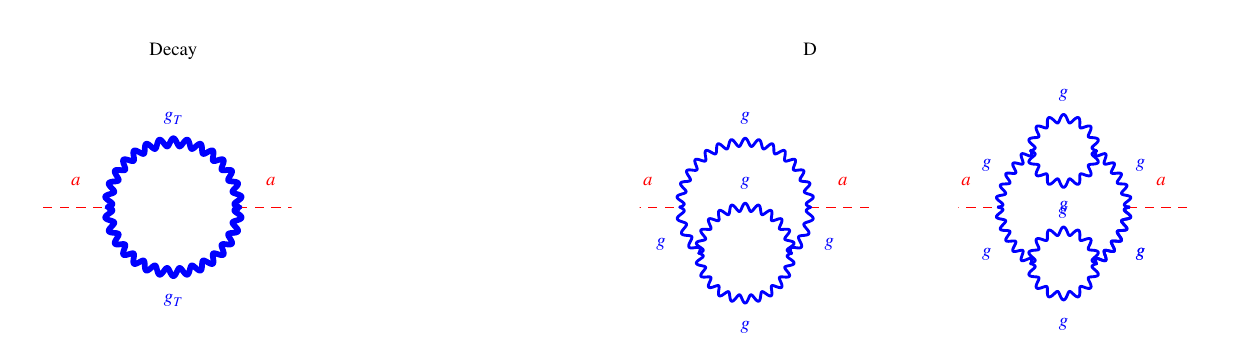}\hspace{0.05\textwidth}
\makebox[0cm][c]{\raisebox{0.106\textwidth}[\height][0pt]{\hspace{-0.7\textwidth}=\hspace{0.32\textwidth}+\hspace{0.21\textwidth} $+\cdots$}}
$$

\caption{\label{fig:FeynDec}\em 
The thermal  diagram `Decay', where the gluon propagator includes one-loop thermal corrections,
is equivalent to the thermal diagram D (thick lines denote the propagator of the thermal gluon $g_T$)
plus the resummation of higher order diagrams with corrections to the gluon propagator.}
\end{center}
\end{figure}

However, both computations give an infra-red divergent result, because of the massless gluon in the $T$ and $U$ diagrams,
or equivalently in the thermal diagram $D$.
We employ the thermal field theory formalism because it is more appropriate for dealing with such issues.

%resumming higher order thermal corrections which regulate the infra-red divergence.

The infra-red divergence is regulated by the thermal gluon mass. 
We re-sum the thermal effects that modify the gluon dispersion relation by substituting the two-loop thermal diagram $D$ with the one-loop `Decay' diagram of fig.\fig{FeynDec}, where the 
tree-level gluon propagator is replaced by the full thermal gluon propagator at leading order in the strong coupling.
In a diagrammatic expansion, the `Decay' diagram corresponds to diagram $D$, plus all higher order diagrams
with any number of corrections to the gluon propagator, as illustrated in the right-handed side of
figure\fig{FeynDec}.
We give the name `decay' to such resummed diagram because physically it describes the decay process $g_T\to g_T a$
of the thermal gluon $g_T$, opened by the non-relativistic thermal corrections to the gluon propagator.
The rationale for re-summing this class of higher-order effects is that they are enhanced
by the $2\to 1$ phase space factor, which is $\sim (4\pi)^2$ bigger than the phase space
relative to $2\to2$ scatterings.

\smallskip

\medskip

%The total scattering rate is the sum of various $2\to2$ processes $\alpha$,
%$\beta$ and $\gamma$, listed in table~\ref{tab:diffcs}. In the present case we have:
%\[
%\gamma = |\alpha_{s} + \alpha_{t}+\alpha_{u}+\alpha_{x}|^{2} +
%|\beta_{s}|^{2} + |\gamma_{t} |^{2},
%\]
%where the labels indicate the $s$-, $t$-, $u$- or $x$-channel.
%The main result can be obtained by careful visual inspection of
%cutting rules; e.g. diagrams of type D (that is a diagram involving gluon propagators corrected by self-energy contributions ) describes the sum $|\alpha_{s}|^{2} +
%| \alpha_{t}|^{2} + | \alpha_{u}|^{2} + | \alpha_{s}|^{2} + | \alpha_{t}|^{2}
%+|\beta_{s}|^{2}+|\gamma_t|^2$ of the modulus squared of all $2\to2$ diagrams
%that contain the gauge coupling $g_{3}$. (Notice that the imaginary part of a single two-loop diagram can
%describe contributions to different scattering processes). An important point is that thermal masses open a phase space for $1\to2$ processes. In order to take this into account we will use the full resummed gluon propagator in the diagram of type D.

In conclusion, the resummed total axion production rate is computed as
\begin{equation}
\gamma=\gamma_A + \gamma_B + \gamma_C + \gamma_{\rm Decay}\equiv 
\gamma_{\mathrm{sub}}+\gamma_{\rm Decay}.
\end{equation}
%the diagram D (that describes decay plus modulus squared of
%many single $2\rightarrow2$ diagrams) plus the set of remaining
%$2\rightarrow2$ rates, obtained by subtracting from the total
%scattering rate $\gamma_S$ the effects already included in
%$\gamma_D$.
The computation of $\gamma_{\rm sub}$ (subtracted scattering rates) is presented in 
 section \ref{subtracted scattering rate}, and the computation of  $\gamma_{\rm Decay}$ is presented in section \ref{Full axion  production rate}.
  Unlike in the HTL approximation,  our technique does not need the
introduction of an arbitrary splitting scale $k_{\ast}$ that satisfies the
problematic conditions $g_3T\ll k_{\ast}\ll T$ in order to control
infra-red divergences. The total rate will be positive for any $g_3$.

While we omitted quarks and other axion couplings to simplify the above discussion,
of course we take them into account  in the full computation.

\section{Subtracted scattering rates}\label{subtracted scattering rate}

Table~\ref{tab:diffcs} lists the full scattering rates and the subtracted scattering
contributions to the various axion production  processes due to the axion/gluon/gluon interaction. It is important to
notice that, unlike the total rate, \emph{the subtracted rates are
infra-red convergent} as expected: the infra-red divergent factors 
$1/t$ and $1/u$ present in the full rate disappear from the subtracted rate.
Actually, by performing computations in the Feynman gauge, we find that $\gamma_{\rm sub}=0$.
The same holds for the other SM vectors.

In order to double-check our result, we computed the subtracted scattering rates also as thermal diagrams
that contribute to the non time-ordered axion propagator $\Pi^<_a$, see eq. (\ref{eq:ImProp}).
This computation is presented in the next part of this section.
 At the end of this section we also evaluate the top Yukawa contribution, which emerges from the axion interaction term in (\ref{lambdatprime}).

%Let us now compute the subtracted scattering rate, $\gamma
%_{\mathrm{sub}}$, given in Table \ref{tab:diffcs}. 
%Since we assume the axion to be weakly coupled,   we will equivalently compute the (subtracted part of the) n. We show here that $\gamma
%_{\mathrm{sub}}=0$.
%

\subsubsection*{Diagram A}
We first consider the contribution of the thermal diagram A in fig.~\ref{fig:FeynA}.  Making use of the Kobes-Semenoff rules (see e.g.~\cite{Salvio:2011sf}) we obtain the following contribution of the diagram A to $\Pi^<$: 
\begin{equation} \Pi^<_A =F\int \frac{d^4K}{(2\pi)^4} \int \frac{d^4Q}{(2\pi)^4} \Delta^<(P-K)  \Delta^<(K-Q) \Delta^<(Q) {\rm Re}\left[\Delta(K) \Delta(P-Q)^*\right]  \alpha(P,K,Q), \label{PiA} \end{equation}
where $F\equiv 2c_i'^2 g_i^6/[3(4\pi)^4f_a^2]$, $P$ is the axion momentum and $\Delta$ and $\Delta^<$ are respectively the tree level scalar propagator and non time-ordered propagator at finite temperature:
\begin{equation} \Delta(K) =\frac{i}{K^2+i\epsilon } + 2\pi n_B(K_0) \delta(K^2) , \qquad \Delta^< (K)= (\theta(-K_0)+n_B(K_0))2\pi \delta(K^2), \label{tree-level-prop}\end{equation}
where $n_B(x)\equiv (\exp(|x|/T)-1)^{-1}$.
These emerge from the scalar part of the gluon propagators, while the contraction of Lorentz and color indices leads to the function $\alpha$ defined by 
\begin{eqnarray}  \alpha(P,K,Q)&\equiv &C_{N}\left[ 4(Q\cdot P)^2(K\cdot P-4K^2)-2 K\cdot P Q\cdot P(K^2+Q^2-14K\cdot Q-2K\cdot P)\right.  \nonumber \\ && \left. -16Q^2 (K\cdot P)^2\right],\end{eqnarray}
where we have introduced $C_{N}\equiv N(N^{2}-1)$ with $N=3,2,1$ for $\SU(3)_c$, $\SU(2)_L$ and U$(1)_Y$ respectively.
%Here  we take the axion mass, $m_a$, to be zero: indeed $m_a$ is proportional to $1/f_a$ and taking into account $m_a$ here would require to go at next to leading order in   $1/f_a$, which goes beyond the scope of this paper (and  beyond the validity of eq. (\ref{eq:ImProp})). 
Looking at the thermal diagram we find that the kinematical configurations with non vanishing phase space in the integrand of eq. (\ref{PiA}) are 
\begin{enumerate}
\item $P_0-K_0 <0, \quad Q_0 >0, \quad K_0-Q_0>0 $;
\item $Q_0<0, \quad P_0-K_0>0, \quad K_0-Q_0>0$;
\item $K_0-Q_0<0 \quad Q_0>0,\quad P_0-K_0>0.$
\end{enumerate}
As anticipated in eq.\eq{ABCD}, these contributions correspond to the interferences between the $s$-channel, $t$-channel and $u$-channel scattering diagrams.
Although  these contributions are separately non-trivial, we find that their sum is identically zero. So
$ \Pi^<_A =0$.

\begin{table}[t]
\begin{center}%
%\begin{tabular}
%[c]{c|r@{~$\to$~}l|ccc}
%& \multicolumn{2}{c|}{process} & $\gamma_{D}$ & $\gamma
%_{\mathrm{sub}}$ & \\\hline\hline $\alpha$ & $g g $ & $g a$
%& $|\alpha_s|^2+|\alpha_t|^2+|\alpha_u|^2$ & $0$ &
%\\\hline
%$\beta$ & $q \bar{q}$ & $g a $ & $|\beta_s|^2$ & $0$ &
%\\\hline $\gamma$ & $q g$ & $q a $ & $|\gamma_t|^2$ &
%$0$ & \\\hline \end{tabular}
%\medskip
%
$$
\begin{array}{ccc}
\hbox{process} & |\mathscr{A}|^2/(g_3^6/128\pi^2f_a^2)& |\mathscr{A}|^2_{\rm sub}/(g_3^6/128\pi^2f_a^2)\\[1mm] \hline
 g g \to  g a & -4|f^{abc}|^2 (s^2+st+t^2)/st(s+t)  & 0  \\
 q \bar{q}\to g a  &  |T^a_{ji}|^2 (s+ 2t + 2 t^2/s ) & 0\\
q g\to q a  &  |T^a_{ji}|^2 (-t-2s - 2 t^2/s ) & 0\\
\end{array}$$
\end{center}
\caption{\textit{Axion production rate from the axion/gluon/gluon interaction.
The total rate is gauge independent; the subtracted rate is gauge dependent and computed in the Feynman gauge.}}%
\label{tab:diffcs}%
\end{table}

\subsubsection*{Diagram B}
We now turn to the thermal diagram B in fig. \ref{fig:FeynA}. We find:
\begin{equation} \Pi^<_B =4F\int \frac{d^4K}{(2\pi)^4} \int \frac{d^4Q}{(2\pi)^4}  \Delta^<(K) \Delta^<(Q) \Delta^<(P-K-Q){\rm Im}\left[\Delta(P-K)\right]  \beta(P,Q,K). \label{PiB} \end{equation}
The function $\beta$, which results from the contraction of Lorentz and color indices, is 
 \begin{equation}\beta(P,Q,K)\equiv 3C_{N}\left[K\cdot P\left(K\cdot P+2K\cdot Q+2Q\cdot P\right)-2K^2Q\cdot P\right].\end{equation}
%
%where again we took the limit $m_a\rightarrow 0$.  
The phase space of the integrand in (\ref{PiB}) can be divided in three regions, which correspond  to the interferences between the $x$-channel and the $s$, $t$ and $u$-channels respectively:
\begin{enumerate}
\item $K_0 <0, \quad Q_0 >0, \quad P_0-K_0-Q_0>0 $;
\item $Q_0<0, \quad K_0>0, \quad P_0-K_0-Q_0>0$;
\item $P_0-K_0-Q_0<0, \quad Q_0>0,\quad K_0>0$.
\end{enumerate}
The sum of the three contributions in the phase space gives zero like for Diagram A: %
$ \Pi^<_B =0$.
% 
%
%$$ \Pi^<_B =\frac{3C_{N}F}{(2\pi)^4}\int dk \, dz_k \, dq \,dz_q \frac{f_B(k)f_B(q)(1+f_B(k+q-p))}{D(k,q,z_k,z_q)^{1/2}}\left[k p(1-z_k)-q p(1-z_q)\right], $$
% 
%where the integral is performed on the intersection between the domain 
%
 %\begin{equation}0\leq p < \infty, \quad 0\leq k < \infty, \quad -1 \leq z_k\leq 1, \quad 0\leq q < \infty, \quad -1 \leq z_q\leq 1\end{equation}
%
%and 
%
 %\begin{equation} D (k,q,z_k,z_q) \equiv (1-z_k)(1-z_q)-\left[1-z_k z_q-\frac{p}{q}(1-z_k)-\frac{p}{k}(1-z_q)\right]^2\geq 0.\end{equation}
 %
 %Since the integrand in $\Pi^<_B$ is antisymmetric in the exchange of $k$ and $q$ we get 
%$ \Pi^<_B =0$.
% It is easy to show that the contribution of $\Pi^<_B$ to the integrated axion thermal production $\gamma_B$ can be written as 
% %
% $$ \gamma_B= \frac{3C_{N_c}}{(2\pi)^6} F \,T^6 b$$
% %
% and $b$ is a number, $b\simeq 23$.

\subsubsection*{Diagram C}

Finally, we evaluate the thermal diagram C in fig.~\ref{fig:FeynA}, the contribution of the $x$-channel alone. It contributes to $\Pi^<$ an amount 
\begin{equation} \Pi^<_C =18FC_{N} \int \frac{d^4K}{(2\pi)^4} \int \frac{d^4Q}{(2\pi)^4}  \Delta^<(K) \Delta^<(Q) \Delta^<(P-K-Q) P\cdot Q.  \end{equation}
Like for  thermal diagrams A and B we can divide the phase space in three parts, which this times correspond to the possibility  to choose one out of three gluons and put it in the final state. The sum of these three contributions vanishes like for diagram A and B:
$ \Pi^<_C =0$.

\medskip

%Therefore, the subtracted part of the scattering rate vanishes, $\gamma_{\rm sub}=0$ 
%so 
%%
%\begin{equation} \gamma_{\rm sub}= \frac{C_{N_c} b \,g_3^6 T^6}{8(2\pi)^{10} f_a^2}, \qquad \mbox{with} \quad  b\simeq 23 \label{gamma-sub} \end{equation}
%%

\subsubsection*{The top Yukawa contribution}
The axion interaction in eq.~(\ref{lambdatprime}) produces the following contribution to $ \Pi^<_a$ at the first non-trivial order in the perturbative expansion
\begin{equation} \Pi^<_t =-24\left(\frac{c_t'y_t }{f_a}\right)^2 \int \frac{d^4K}{(2\pi)^4} \int \frac{d^4Q}{(2\pi)^4}  Q\cdot  P\Delta_F^<(K)\Delta_F^<(P-K-Q) \Delta^<(Q),
 \end{equation}
where $\Delta^<_F$ is the tree level non time-ordered propagator at finite temperature for a fermion:
\begin{equation} \Delta_F^<(K) \equiv  [\theta(-K_0)-n_F(K_0)]2\pi \delta(K^2) \label{deltaF}
 \end{equation}
and $n_F(x)\equiv (\exp(|x|/T)+1)^{-1}$. Like in the previous computation, the integral receives contributions from three
distinct integration regions, which correspond to the effects that can be equivalently computed as
$Q_3 H^* \rightarrow a \bar U_3$, $U_3 H^* \rightarrow a \bar Q_3$ and $Q_3 U_3 \rightarrow a H$ scatterings
(as well as their CP-conjugated processes):
 \begin{enumerate}
\item $K_0 <0, \quad P_0-K_0-Q_0 >0, \quad Q_0>0 $;
\item $K_0>0, \quad  P_0-K_0-Q_0<0, \quad  Q_0>0$;
\item $K_0>0, \quad P_0-K_0-Q_0 >0,\quad  Q_0<0$.
\end{enumerate}
The first possibility, for example, leads to the following contribution to the production rate
\begin{equation} \gamma_{t1} =  \frac{6c_t'^2y_t ^2}{(2\pi)^6 f_a^2}\int dp\, dk \, dz_k \, dq \,dz_q p^2 q (1-z_q)\frac{(1-n_F(k))n_F(p+k-q) n_B(q)}{D_1(k,q,z_k,z_q)^{1/2}}, \end{equation}
where the integral is performed on the intersection between the domains 
 \begin{equation} 0\leq p < \infty, \quad 0\leq k < \infty, \quad -1 \leq z_k\leq 1, \quad 0\leq q < \infty, \quad -1 \leq z_q\leq 1\end{equation}
and 
 \begin{equation} D_1 (k,q,z_k,z_q) \equiv (1-z_k)(1-z_q)-\left[-1-z_k z_q+\frac{p}{q}(1+z_k)-\frac{p}{k}(1-z_q)\right]^2\geq 0.\end{equation}
 These conditions emerge because $k$ and $q$ are the lengths of the three dimensional parts ($\vec{k}$ and  $\vec{q}$) of the on shell momenta $K$ and $Q$, once the delta functions in (\ref{deltaF}) are used, and $z_k$ and $z_q$ are the cosines of the angles between $\vec{k}$ and $\vec{p}$ and $\vec{q}$ and $\vec{p}$ respectively.

 The other two contributions lead to similar expressions. The total result due to the  interaction in (\ref{lambdatprime}) is 
  \begin{equation}
 \gamma_a^{\rm top}= 0.94 \frac{3 y_t ^2 c_t^{\prime2} T^6}{2\pi^5 f_a^2},
% \frac{3y_t ^2 c^{\prime 2}_t T^6}{(2\pi)^6 f_a^2} k_t, \qquad \mbox{where} \quad k_t\simeq 47. \label{resultt}
\end{equation}
where the numerical factor is the Bose-Einstein and Fermi-Dirac
correction with respect to the analytic result computed in Boltzmann approximation, $n_{B,F}(E) \approx e^{-E/T}$.

\setcounter{equation}{0}
\section{Thermal vector decays}\label{Full axion  production rate}

We start summarizing some well known results from quantum field theory at finite temperature
that are relevant for our computations.

%\begin{figure}[t]
%\begin{center}
%$$\includegraphics[width=\textwidth]{fSUSYT}$$
%\caption{\label{fig:SUSYMT}\em Dispersion relations at finite temperature in HTL approximation
%for the components within a chiral (left) and vector (right) massless super-multiplet. Thermal effects are supersymmetric at $k\gg m\sim gT$.}
%\end{center}
%\end{figure}

\subsection{Thermal corrections to  vector propagators}
%The HTL approximation~\cite{Pisarski92,LeBellac} holds at momenta $k$ and energies $\omega$
%that satisfy $k,\omega \ll T$,
% describing thermal effects that arise at $k,\omega \sim gT$ if $g\ll 1$.
%In HTL approximation, the  key parameters is the  thermal mass $m$.
%It means that the dispersion relation of transverse vector polarizations at large momentum satisfies
%$\omega^2(k\gg T) \simeq k^2 + m^2$, while at rest both transverse and
%longitudinal polarizations have energy $\omega^2(k=0)=2 m^2/3$.
%
% 
% 
%However, the physically relevant value of the strong gauge coupling
%is not small enough to justify the use of the HTL approximation.
%We therefore use the full one-loop thermal and quantum corrections to the vector propagator~\cite{WeldonFermion,WeldonVector,thoma,LeBellac}.
%

 \label{Vector propagator}

We list the full one-loop expressions for thermal corrections to a vector~\cite{WeldonVector,thoma,LeBellac}
with four-momentum $K=(\omega,\vec{k})$ ($K^2=\omega^2-k^2$)
with respect to the  rest frame of the thermal plasma.
We denote by $U_\mu$ the four-velocity $U_\mu$ of the plasma.
We use the Feynman gauge where all effects are condensed in two
form factors even in the non-abelian case~\cite{WeldonVector}.
Polarizations are conveniently decomposed in
 $T$ransverse
(i.e.\ orthogonal to $K$ and to $\vec{k}$), $L$ongitudinal
(i.e.\ orthogonal to $K$ and parallel to $\vec{k}$) and
parallel to $K$.
The corresponding projectors $(\Pi^T +\Pi^L +\Pi^K)_{\mu \nu}=-\eta_{\mu\nu}$ are
\begin{eqnsystem}{sys:PiV}
\Pi_{\mu\nu}^T &=&-\tilde{\eta}_{\mu\nu}+\frac{\tilde{K}_\mu\tilde{K}_\nu}{-k^2}=
 \begin{pmatrix}
0 & 0\cr 0 & \delta_{ij}- k_i k_j/k^2 \end{pmatrix},\\
 \Pi_{\mu\nu}^L &=& -\eta_{\mu\nu} + \frac{K_\mu K_\nu}{K^2} -\Pi_{\mu\nu}^T ,\\
\Pi_{\mu\nu}^K  &=&  - \frac{K_\mu K_\nu}{K^2},
\end{eqnsystem}
where $\tilde{\eta}_{\mu\nu}=\eta_{\mu\nu}-U_\mu U_\nu$, $\tilde{K}_{\mu} = K_\mu - (K\cdot U) U_\mu$.
The vector propagator is~\cite{WeldonVector,thoma}
\beq^*D_{\mu\nu} =  i\bigg[\frac{\Pi_{\mu\nu}^T}{K^2-\pi_0- \pi_T} + \frac{\Pi_{\mu\nu}^L}{K^2-\pi_0-\pi_L} +  \frac{\Pi_{\mu\nu}^K}{K^2}\bigg]\eeq
where the corrections are contained in the scalar functions $\pi_0(k,\omega)$
(quantum corrections at $T=0$) and
 $\pi_T(k,\omega)$ and $\pi_L(k,\omega)$ (thermal corrections),
explicitly given in~\cite{Rychkov:2007uq} for a general theory.
The corresponding non-time ordered propagator is
\begin{equation} ^*D_{\mu\nu}^<(K) = f_B(k_0)\bigg[ \Pi_{\mu\nu}^T \rho_T(K)+\Pi_{\mu\nu}^L \frac{|\mb{k}|^2}{K^2}\rho_L(K)+
\xi \frac{k_\mu k_\nu}{K^4}\bigg].\label{eq:rhoV}\end{equation}
Here,  $\rho_T$, $\rho_L$ are the spectral densities for
the  transverse vectors and longitudinal vectors respectively
\beq \rho_T = -2\Im  \frac{1}{K^2 -\pi_0-\pi_T},\qquad \rho_L
=-2\Im \frac{K^2}{k^2}\frac{1}{K^2-\pi_0-\pi_L}. \eeq 
Furthermore, $k_0>0$ describes a  vector in the final state,
and $k_0<0$ describes a  vector in the initial state:
this convention allows to compactly describe all possible processes. Indeed
the factors
%\begin{equation}f_B(k_0)\equiv \frac{1}{e^{k_0/T}-1} = \left\{\begin{array}{ll}
%n_B & \hbox{if $k_0>0$} \end{equation}
\beq
f_B(k_0)\equiv \frac{1}{e^{k_0/T}-1} = \left\{\begin{array}{ll}
n_B(k_0) & \hbox{if $k_0>0$}\\
-(1+n_B(k_0)) & \hbox{if $k_0<0$}
\end{array}\right.
\eeq
gives the usual statistical factors: $-n$ (number of particles in the initial state) or $1\pm n$ (stimulated emission or Pauli-blocking in the final state),
where $n_{B}(E)\equiv 1/(e^{|E|/T}- 1)$ is the Bose-Einstein distribution. 

\medskip  
\begin{figure}[t]
\begin{center}
$$\includegraphics[width=0.7\textwidth]{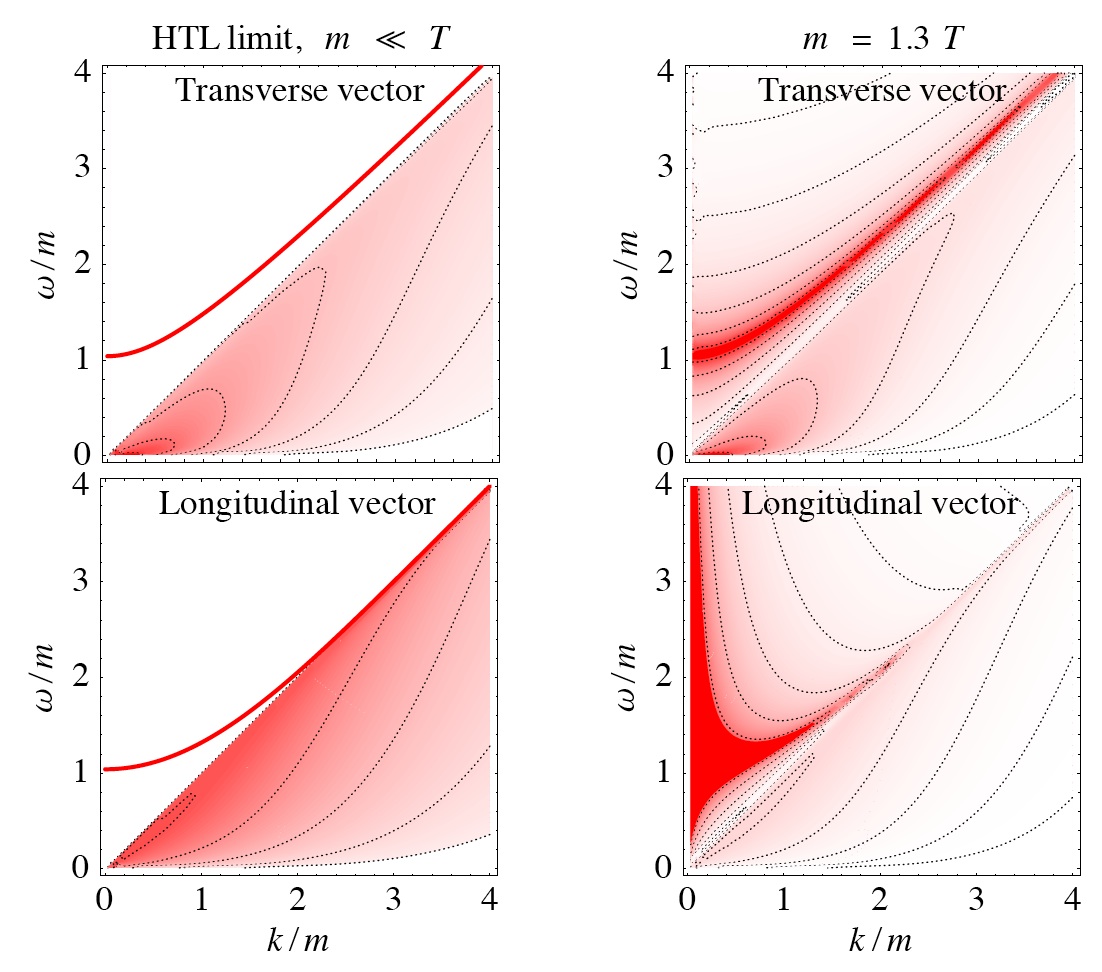}$$
\caption{\label{fig:Vrho}\em 
One loop thermal densities $\rho_T(\omega, k)$ (upper row) and $\rho_L(\omega, k)$ (lower row)
for a vector in the Hard Thermal Loop limit $g\ll1$ (left) and for $g\sim 1$ (right).
In the HTL limit there is a pole above the light cone and a continuum below the light cone.
In the full case the pole above the light cone acquires a finite width and becomes a continuum, and
 the continuum below the light cone gets Boltzman suppressed at $k\gg m$. }
\end{center}
\end{figure}

Fig.\fig{Vrho} shows the spectral densities in the HTL limit ($g\ll1$, left panels)
and in a realistic situation ($g\sim 1$, right panels).
In the realistic case, the poles get smeared acquiring a finite width.
A more significant difference arises below the light cone ($\omega<k$),
where both the HTL and the one-loop spectral densities do not vanish. 
This
describes `Landau damping' i.e.\ the fact that particles  exchange
energy with the thermal plasma. 
However, the HTL approximation cannot
be applied at $k\sim T$ (a region relevant for us, since $g\sim 1$),
and indeed it misses that at $k \gg T$
spectral densities get suppressed by an exponential Boltzmann
factor. 
%This Boltzmann suppression of the spectral density below the light cone
%is the main difference between  the HTL approximation
%and the full one loop.

\subsection{Axion production via vector thermal decays}
In section \ref{subtracted scattering rate} we have computed the subtracted axion production rate; 
we here compute the resummed `Decay' diagram of fig.\fig{FeynDec}, which reduces at leading order to diagram D in fig.~\ref{fig:FeynA}. 
As already stated, the rationale for re-summing only this class of higher-order
effects is the phase space enhancement of the 2 $\rightarrow 1$ processes (relative to the $2 \rightarrow 2$ scatterings). Therefore, the residual gauge dependence in our result
is expected to be of relative order $g^2/\pi^2$.
The computation applies to all SM vectors $V=\{g, W, Y\}$ with gauge couplings $\alpha_i = \{ \alpha_3, \alpha_2,\alpha_Y\}$ and
dimension of the gauge group $d_i =\{8,3,1\}$.

The resummed contribution to the axion propagator $\Pi^<_a$ is 
\beq  \Pi_{\rm res}^< = \frac{c_i'^2}{f_a^2}  \frac{d_i\alpha_i^2}{8 \pi^2} \int \frac{d^4 K}{(2\pi)^4}
\epsilon^{\mu \nu \alpha \beta } \epsilon^{\mu' \nu' \alpha' \beta'} K_\alpha Q_\beta K_{\alpha'} 
Q_{\beta'}  \   ^*D_{\mu\mu'}^<(K) ^*D_{\nu\nu'}^<(Q),
\eeq
The vector quadri-momenta are $K_\mu$ and $Q_{\mu}=P_\mu- K_\mu$, where $P_\mu$ is the axion quadri-momentum. Inserting the parametrization
\beq P= ( p_0, p,0,0)\ , \quad K= ( k_0, k \cos \theta_k. k \sin \theta_k,0) \ ,\quad Q = ( q_0, q\cos \theta_q , q \sin \theta_q ,0) ,\eeq
we obtain
\begin{eqnarray}
\Pi_{\rm res}^< &=& \frac{c_i'^2}{f_a^2}  \frac{d_i\alpha_i^2}{8 \pi^2} \int \frac{d^4 \!K}{(2\pi)^4} f_B(k_0) f_B(q_0)  \times
  \nonumber\\ &&  \bigg\{ \left( \rho_L(K) \rho_T( Q) + \rho_T(K) \rho_L(Q) \right)  k^2 q^2 \sin^2( \theta_k -\theta_q )   + \nonumber\\ &&  \rho_T(K) \rho_T( Q) \left[  \left(k_0^2 q^2 + k^2 q_0^2\right) \left(1+ \cos^2( \theta_k -\theta_q )\right)  - 4 k_0q_0 k q \cos ( \theta_k - \theta_q ) 
 \right] 
\bigg\}  ,
\end{eqnarray}
where we used the decomposition of the resummed propagator given in (\ref{eq:rhoV}). In order to compute the integral, it is convenient to multiply by $1 = \int d^4Q \,\delta \left( K+Q-P\right)$. After performing the angular integrations over $\theta_k$ and $\theta_q$, and using the equations,
\beq
\cos \theta_q = \frac{-k^2 + q^2 + p^2}{2 pq} ,\quad  \cos\theta_k = \frac{k^2- q^2  + p^2 }{2 kp},\quad \cos\left( \theta_k - \theta_q\right) = \frac{-k^2 - q^2 + p^2}{2 kq}
\eeq
we obtain
\begin{eqnarray}
\Pi_{\rm res}^< &=& \frac{c_i'^2}{f_a^2}  \frac{d_i\alpha_i^2}{8 (2\pi)^5}\frac{1}{p} \int^{\infty}_{-\infty} d k_0 \int^{\infty}_0 d k  \int^{k+p}_{|k-p|} d q \ kq  f_B(k_0) f_B(p_0 - k_0)  \nonumber\\
&& \bigg\{ \left( \rho_L(K) \rho_T( Q) + \rho_T(K) \rho_L(Q) \right) \left[  (k+q)^2 -p^2 \right]
 \left[ p^2 - (k-q)^2 \right] + \\
&&  \rho_T(K) \rho_T( Q) \left[ \left( \frac{k_0^2} {k^2} + \frac{ q_0^2}{q^2}\right) 
\left( ( k^2 - p^2 + q^2  )^2 + 4 k^2 q^2 \right) +  8 k_0 q_0  \left( k^2 + q^2 - p^2 \right) \right] \bigg\}.   \nonumber
\end{eqnarray}
Note that the integration range is restricted to $ |p-k|\leq q\leq p+k $.
Finally, for each factor of the SM gauge group we  computed this integral numerically using the spectral densities described in the previous section.
We followed the method provided in \cite{Rychkov:2007uq}. 

\setcounter{equation}{0}

\begin{table}[t]
\begin{center}
$$\begin{array}{c|cccc} \\
\hbox{Gauge group}  &  N &N_F & N_S  &  \hbox{Thermal mass}^2\\ \hline
\hbox{Color $\SU(3)_c$} & 3 &6 &0  & g_3^2 T^2\\
\hbox{Weak $\SU(2)_L$}& 2 & 6 &1/2 &  11 g_2^2 T^2 /12\\
\hbox{Hypercharge ${\rm U}(1)_Y$} &0 & 10 & 1/2 &11 g^2_Y T^2/12 \end{array}$$
\end{center}
\caption{\label{tab:NNN}\em Numerical coefficients for vector thermal mass $m_i^2 = \frac{1}{6}g_i^2 T^2 (N+N_S + N_F/2)$ in the SM in terms of the $\SU(N)$ factor, of
the number of fermions $N_F$ and of scalars $N_S$.}
\end{table}%

\section{Result}\label{res}
\begin{figure}[t]
\begin{center}
$$\includegraphics[width=0.7\textwidth]{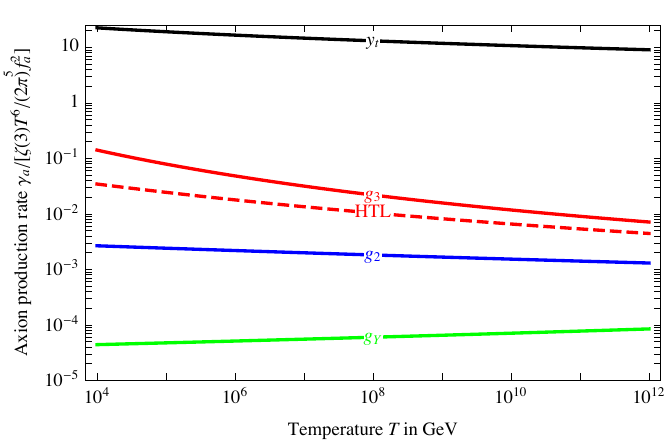}$$
\caption{\label{fig:gamma}\em 
The four contributions to the thermal axion production rate $\gamma_a$
induced by the SM couplings $y_t$ (upper black curve), $g_3$ (red curve),
$g_2$ (blue), $g_Y$ (green) for unity values of the axion couplings $c'_t=c'_3=c'_2=c'_1=1$  in eq.~(\ref{defF321}). 
The red dashed line is the previous result for the strong coupling contribution computed in Hard Thermal Loop approximation.
}
\end{center}
\end{figure}

The total axion production rate due to all axion couplings $c'_3, c'_2, c'_1, c'_t$ and taking into account
all large SM couplings, $g_3, g_2, g_Y, y_t$ is
\begin{equation}
\gamma_{a}=\frac{ T^6 \zeta(3)  }{(2 \pi)^5 f_a^2}\bigg[37 c_t^{\prime2} y_t ^2+
8 c_3'^2\alpha_3^2 F_3(\frac{m_3}{T}) + 3 c_2'^2 \alpha_2^2 F_2(\frac{m_2}{T})
+ c_1'^2 \alpha_Y^2 F_1(\frac{m_1}{T}) 
  \bigg],
 \label{defF321}
\end{equation}
where the thermal masses of SM gauge bosons of $\SU(3)_c$, $\SU(2)_L$ and ${\rm U}(1)_Y$ are
(see table~\ref{tab:NNN}):
\beq
 \frac{m_3}{T}=g_3,\qquad
\frac{m_2}{T}=\sqrt{\frac{11}{12}} g_2,\quad
\frac{m_1}{T}= \sqrt{\frac{11}{12}} g_Y.\eeq
Fig.\fig{res} shows our results for the functions $F_{1,2,3}$ that parameterize the axion production rate 
due to gauge interactions, while our result for the top Yukawa part is given analytically.
%, apart from the numerical overall factor $\simeq 37$.
Fig.\fig{gamma} shows the four contributions to the axion production rate as function of the temperature.
As long as $c'_t \sim c'_3$ the top Yukawa axion production rate gives the dominant contribution
because it arises at tree level, while the anomalous axion couplings arise at loop level.

%$\gamma_3$, due to the axion/gluon coupling, as
%\beq\gamma_3=\frac{c_3'^2\zeta(3)g_3^4T^6}{64\pi^7f_a^2} F_3(g_3) \label{F-def}\eeq
Previous works ignored the top Yukawa effect and
computed only the function $F_3$ within the HTL approximation i.e.\
in the limit of small strong gauge coupling, $g_3\ll 1$.
We see that  in this limit our improved computation reproduces to the HTL limit.
However, when $g_3$ is set to its physical value, $g_3\approx 1$,
the results differ: the HTL approximation breaks down and the HTL rate function $F_3^{\rm HTL}$
becomes unphysically  negative for a large enough $g_3$, while our result grows with increasing $g_3$.

% 
%The previous calculation of axion production due to gluons can  be easily extended to the production due to the electroweak interactions and the top Yukawa, eq. (\ref{lambdatprime}), for temperatures $T\gg M_t$, where the electroweak symmetry breaking can be neglected. Denoting with $\gamma_3$, $\gamma_2$ and $\gamma_1$ the axion production rates due to the $SU(3)_c$, $SU(2)_L$ and $U(1)_Y$ gauge fields respectively we have 
%\begin{equation}
%\gamma_3=\frac{c_3'^2d(SU(3)) \alpha_3^2 T^6 \zeta(3) F_3 }{(2 \pi)^5 f_a^2}, \quad \gamma_2= \frac{c_2'^2 d(SU(2)) \alpha_2^2 T^6 \zeta(3)F_2 }{(2 \pi)^5 f_a^2},  \quad \gamma_1= \frac{c_1'^2 \alpha_Y^2 T^6 \zeta(3)F_1 }{(2 \pi)^5 f_a^2}, \label{defF321}
%\end{equation}
%while the top Yukawa contribution that we find is 
%\begin{equation}
%\gamma_t= \frac{y_t '^2T^6}{(2\pi)^6 f_a^2} k_t, \qquad \mbox{where}\quad k_t \simeq 47   \label{topcontr}.
%\end{equation}

%The rate functions $F_3$, $F_2$ and $F_1$ are functions of $g_3$, $g_2$ and $g_Y$ respectively and in order to compute them one has to take into account the different representations of the SM fields under  
%$SU(3)_c$, $SU(2)_L$ and $U(1)_Y$, summarized in Table \ref{tab:NNN}. 
%%Note that $\gamma=\gamma_3$ and $F_3=F$, defined in eq. (\ref{F-def}).
%
%
%In figure \ref{fig:res} we give $F_3$, $F_2$ and $F_1$ as functions of the corresponding thermal mass, $m$: the gluon, weak and hyper charge thermal mass respectively.

\subsection{Cosmological axion yield}\label{cosmo}
In the usual scenario of reheating after inflation,  the inflaton $\phi$ with energy density $\rho_\phi$
decays with width $\Gamma_\phi$
into SM particles  (excluding the axion).
The reheating temperature $T_{\rm RH}$ is defined as the temperature
at which $\Gamma_\phi$ equals $H_R$, the expansion rate due to the radiation density only:
\beq 
T_{\rm RH}= \left[ \frac{45}{4\pi^3 g_*}\,\Gamma_\phi^2
M_{\rm Pl}^2\right]^{1/4} .\eeq
Here we neglect possible non-equilibrium effects at $T\gg T_{\rm RH}$  \cite{Mazumdar:2013gya}. 
$T_{\rm RH}$ effectively is the maximal temperature of the universe.
Indeed, while higher temperatures exist, particles produced
at $T>T_{\rm RH}$ are diluted by the entropy released by inflaton decays,
as described by the 
$Z - 1 =- {\Gamma_\phi \rho_\phi}/{4 H \rho_R}$ term in the Boltzmann equations
\begin{equation}\label{eq:dY/dz}
\left\{\begin{array}{rcl}\displaystyle
  HZz\frac{d\rho_\phi}{dz}&=& -
{3H\rho_\phi}-{\Gamma_\phi\rho_\phi}\,   ,\\[4mm]
\displaystyle
sHZz \frac{dY_a}{dz} &=& 3sH(Z-1)Y_a+\displaystyle \gamma_a (1- \frac{Y_a}{Y_a^{\rm eq}}).
\end{array}\right.\eeq
Here $H = \sqrt{8\pi\rho/3}/M_{\rm Pl}$ is the Hubble parameter, $z=T_{\rm RH}/T$,
$s=2 T^3 g_{\rm SM} \pi^2/45$ is the entropy density of SM particles
($g_{\rm SM}=427/4$),
$n_a$ is the axion number density,
$Y_a=n_a/s$,  and $Y_a^{\rm eq}=n_a^{\rm eq} /s\approx 0.00258$ with 
$n_a^{\rm eq} = \zeta(3) T^3/\pi^2$
is the thermal equilibrium value of $Y_a$.
The solution to the Boltzmann equations for the  axion abundance at $T\ll T_{\rm RH}$ is
\beq \frac{Y_a}{ Y_a^{\rm eq} } =(1+r^{-3/2})^{-2/3} \simeq
 \left\{ \begin{array}{ll}
r &\hbox{for $r\ll1$}\\
1 & \hbox{for $r\gg 1$}
\end{array}\right.\label{eq:rrr}
\eeq
where
\beq \label{eq:r}
r= \frac{2.4}{Y_a^{\rm eq} }\left.
\frac{\gamma_a}{Hs}\right|_{T=T_{\rm RH}} = 1.7 \frac{T_{\rm RH}}{10^{7}\GeV} \left(\frac{10^{11}\GeV}{f_a}\right)^2
\left. \frac{\gamma_a}{T^6 \zeta(3)/(2\pi)^5 f_a^2} \right|_{T=T_{\rm RH}} .
\eeq
The latter factor in eq.\eq{r} is the order-one term among square brackets in eq.~(\ref{defF321}).
The approximated analytical expression of eq.\eq{rrr} valid for intermediate values of $r$ is obtained by fitting 
the numerical solution.
% CODE USED Leptogenesi/LeptogenesisFast.nb

Even for the lowest possible value of $f_a\circa{>} 5 \times 10^{9}\GeV$ and taking into account the
new top effect in $\gamma_a$, eq.\eq{r} implies that axions decoupled at $T \circa{>} M_Z$,
when the number of relativistic SM degrees of freedom $g_{\rm SM}$ still included all SM particles. 
Thereby, when  SM particles later become non-relativistic, they annihilated heating photons and neutrinos, but not axions.
This means that today, and at the epoch of CMB decoupling,
thermal axions constitute a small fraction of the total relativistic energy fraction, conveniently parameterised by the usual
``effective number of neutrinos'' as\footnote{Today photons have temperature $T_\gamma$ and neutrinos
have temperature $T_\nu = T_0 (4/11)^{1/3}$,
for a total of $g_{*s}=43/11$ effective entropy degrees of freedom.
Axions went out of thermal equilibrium at $T\gg M_Z$ would have a present temperature
$T_a = T_\gamma (g_{*s}/g_{\rm SM})^{1/3} = 0.903\,{\rm K}$,
which corresponds to
$\Delta N_\nu^{\rm eff} = 4({T_a}/{T_\nu})^4/7\approx 0.0264$.
We recall that the SM alone predicts $N_\nu^{\rm eff}\approx 3.046$ where
the small deviation from 3 is due to imperfect neutrino decoupling when electrons become non-relativistic~\cite{Dolgov}.}
\beq \Delta N_\nu^{\rm eff} = 0.0264 \frac{Y_a}{Y_a^{\rm eq}} . \eeq
The phenomenological manifestations in cosmology of a thermal axion component of the universe
are analogous to having an extra freely-streaming neutrino component.
Such effects can be parameterised by the axion contribution to
effective number of neutrinos $\Delta N_\nu^{\rm eff}$ and by the
axion mass $m_a\approx 0.6\,{\rm meV} (10^{10}\GeV/f_a)$. 

Full cosmological bounds in the plane $(\Delta N_\nu^{\rm eff}, m)$ were computed in fig.~6a of~\cite{CS} and, more recently, in fig.~28 of~\cite{Planck}.
In practice, present global fits of cosmological data find $\Delta N_\nu^{\rm eff} =0.48\pm0.48$~\cite{Planck}:
the uncertainty is more than one order of magnitude above the maximal thermal axion effect.
Future experiments which are being discussed, such as CMBpol, can reduce the uncertainty on $N_\nu^{\rm eft}$ to $\pm0.044$~\cite{Melc}.

\setcounter{equation}{0}
\section{Conclusions}\label{concl}
In conclusion, we improved over previous computations of the thermal axion density in two ways:
\begin{enumerate}

\item By including higher-order effect enhanced by the thermal decay kinematics gluon $\to$ gluon + axion.
Unlike the leading order result, which becomes negative for $g_3>1.5$, our result behaves physically
for all relevant values of the strong gauge coupling.

\item By considering all axion couplings to  SM particles; not only to gluons.
The strong interaction contribution receives new contributions from the axion couplings to quarks,
as encoded in the difference between $c_3$ and $c'_3$, eq.~(\ref{cred}).
Electroweak effects are small. 
More importantly, as long as the axion couples to the top quark,
there is a new effect related to the top Yukawa coupling, which dominates by 3 order
of magnitude over the effect related to the strong gauge coupling (see fig.\fig{gamma}).

\end{enumerate}
Our result for the thermal axion production rate is given in eq.~(\ref{defF321}) in terms of the 
axion couplings
$c'_3$ (strong interactions), $c'_2$ (weak interactions), $c'_1$ (hypercharge) and
$c'_t$ (top Yukawa coupling) defined in eq.~(\ref{cred}) and~(\ref{lambdatprime}).
The thermal functions $F_3, F_2,F_1$ are numerically plotted in fig.\fig{res}.

Furthermore, we have shown that there are no collinear enhancements (in analogous computations
such effects are present and require resummation of extra classes of thermal diagrams).

The thermal axion abundance is then computed  adopting the usual simplified model of reheating
(rather than the instantaneous reheating approximation adopted in previous works) and allowing for the possibility of a thermalised axion.
We provide in eq.\eq{rrr} a simple numerical  approximation for the final axion abundance.

\small

\paragraph{Acknowledgements}
We thank Simon Caron-Huot, Gian Giudice, Julien Lesgourgues, Alessandro Melchiorri,
for useful discussions and in particular Slava Rychkov for having clarified the physics of collinear enhancements.
We thank Daniel Green and Benjamin Wallisch for having pointed out a numerical error in the electroweak gauge contribution in fig.~\ref{fig:gamma}, which has been corrected.
This work was supported in part by the European Programme PITN-GA-2009-23792 (UNILHC). 
The work of Alberto Salvio has been supported by the Spanish Ministry of Economy and Competitiveness %under grant FPA2012-32828, 
Consolider-CPAN (CSD2007-00042), the grant  SEV-2012-0249 of the ``Centro de Excelencia Severo Ochoa'' Programme and the grant  HEPHACOS-S2009/ESP1473 from the C.A. de Madrid. This work was supported by the ESF grant MTT8 and by
SF0690030s09 project.

%\newpage

\footnotesize

%\bigskip\bigskip

\begin{multicols}{2}

\end{multicols}

\end{document}